\documentclass[aps,prl,reprint,groupedaddress,nofootinbib]{revtex4-2}

\usepackage{graphicx}
\usepackage{xcolor}
\usepackage{amsmath}
\usepackage{amsfonts}
\usepackage{comment}
 \usepackage[normalem]{ulem}

\newcommand{\mn}{{\mu\nu}} 

\begin{document}


\title{Relativistic Fractons and their Dust}



\author{Qiuyue Liang}
\email{qiuyue.liang@ipmu.jp}
\author{Tom Melia}
\email{tom.melia@ipmu.jp}

\affiliation{Kavli Institute for the Physics and Mathematics of the Universe (WPI), University of Tokyo,
Kashiwa, Chiba 277-8583, Japan}  


\date{\today}

\begin{abstract}
We define a relativistic version of the global symmetries responsible for the restricted mobility of fracton quasiparticles. The theories have a symmetry current that is proportional to a vector field that spontaneously breaks Lorentz boost symmetry. We argue that the existence of a pressureless dust in the early universe could be a consequence of this symmetry. 
We provide an example of a fractonic scalar field with a quartic self-interaction evolving on a Friedmann-Robertson-Walker background and show that the interaction gives rise to a separately conserved fluid with equation of state $w=1$. 
\end{abstract}


\maketitle


\noindent\textit{Introduction:} 
 Of all the generalizations of the notion of symmetry in quantum field theory that have been  developed in recent years~\cite{McGreevy:2022oyu, Brennan:2023mmt}, it is perhaps the fractonic symmetries~\cite{Chamon:2004lew,Haah:2011drr,Vijay:2015mka,Vijay:2016phm} that give rise to the most exotic phenomena. Theories with a fractonic symmetry exhibit quasiparticle excitations---fractons---that have highly restricted or no mobility (for reviews, see~\cite{Nandkishore:2018sel,Pretko:2020cko}). 
The restricted motion has been understood using non-relativistic one-form symmetries~\cite{Seiberg:2019vrp}, and dipole and higher multipole symmetries~\cite{Pretko:2018jbi}. Heuristically, a fractonic charge cannot move if there is a global symmetry that  causes the overall dipole moment of the system to be conserved, because such individual motion would change the dipole---at best charges can only move in pairs.
The inclusion of higher multipole symmetries further restricts their possible motion, which is said to become {\it fractionalized}, hence the name {\it fracton};  total immobility is a result of an infinite number of space-dependent symmetries.

Fractonic symmetries are inherently non-relativistic, as the dipole, quadrupole \textit{etc.} symmetries must be defined in a specific frame. However, this can always be viewed from a relativistic perspective, where the Lorentz symmetry has been spontaneously broken \cite{Nicolis:2015sra,Nicolis:2013lma}. In this letter, we define a relativistic version of
fractonic symmetries by defining the symmetry with a vector field that spontaneously breaks Lorentz boosts.  The simplest example of such a field is a `framid'~\cite{Nicolis:2015sra}, and we shall use this setup when we consider a specific model; see~\cite{Jain:2024ngx} for an approach to making multipole symmetries relativistic by constructing a reference solid.

We show that there is an immediate consequence of relativistic fractonic symmetries in the homogeneous and isotropic universe: they enforce the existence of a covariantly conserved, pressureless dust component in the Einstein equations, no matter what interactions the fields on which they act may have. The effect of the interactions instead shows up as a secondary homogeneous fluid that is decoupled from this `fractonic dust'. The dust  could play a role as the dark matter of the universe. 

The breaking of Lorentz boost symmetry by a framid has been well-studied in a cosmological setting, under the name of Einstein-Aether theories~\cite{Gasperini:1987nq,Jacobson:2000xp}. 
Here, the background solution for the framid on a homogeneous and isotropic Friedmann-Robertson-Walker (FRW) metric acts to rescale Newton's constant, with the effect of slowing down the expansion of the universe~\cite{Carroll:2004ai}. The perturbations around this background  have also been studied~\cite{Jacobson:2004ts,Lim:2004js}. 
Using a framid to define a fractonic symmetry, however, results in a fracton density becoming pinned to it, and thus to very different dynamics; in a cosmological application the resulting system could represent a novel coupling between an aether and dark matter.

We proceed by first defining the relativistic version of a theory with a fractonic symmetry, and we show how it implies a pressureless fractonic dust component in the homogeneous and isotropic universe. 
We then turn to a specific example of a fractonic scalar field that has a quartic self-interaction. We first study the fractonic dust that arises in this theory in flat space to discern the low energy dynamics of the coupled fractonic scalar field and the framid. We then show that on a FRW background the energy-momentum tensor consists of a dark matter component and a separate fluid with an equation of state $w=1$ arising from a quartic self-interaction. We conclude with a short discussion.
\newline

\noindent\textit{Relativistic Fractonic Symmetry:} 
Let us first consider the non-relativistic story, in order to see how it can be made relativistic. We start with a theory that has a conserved  current that takes the form 
\begin{eqnarray}
    J^\mu = \rho_c \,\delta^{\mu}_0\,,
    \label{eq:fullf}
\end{eqnarray} 
where $\rho_c$ is to be interpreted as a charge density. Because the current is conserved, $\partial_\mu J^\mu=0$, this implies
\begin{eqnarray}
    \partial_\mu J^\mu = \partial_0 \rho_c = 0 \,.
\label{eq:consnonrel}
\end{eqnarray}
The charge density is, therefore, only a function  of spatial coordinates, $\rho_c=\rho_c(\vec{x})$. This implies an infinite number of conserved charges,  beyond the usual charge of the above current: for any arbitrary function $C(\vec{x})$ of the spatial coordinates, a corresponding conserved charge can be constructed as
\begin{eqnarray}
    Q_{C} = \int d^3\vec x \,C(\vec{x})\, J^0  \,,~~~\partial_0 Q_c=0\,.
\end{eqnarray}
Note that, \textit{e.g.} the usual conservation of total charge corresponds to $C(\vec{x})=1$, a conserved dipole moment to $C(\vec{x})=x^i$, \textit{etc.}.

We can relativisticise the current in Eq.~\eqref{eq:fullf}  by introducing a time-like vector field $u^\mu$ that has a fixed norm, $u^\mu u_\mu=-M^2$. This $u^\mu$ field is taken to be a dynamical field of the theory that spontaneously breaks the (local, when coupling to gravity) Lorentz boost symmetries at the mass scale $M$, and thus provides a preferred coordinate system for the fractonic symmetries to be defined in. {Our viewpoint here is more broad than that in the latter sections where we specialize to $u^\mu$  being a framid (or than that in Ref.~\cite{Jain:2024ngx} where it is defined relative to crystal coordinates).} 

Given such a field, we can look for  theories that have a conserved  current that takes the form
\begin{eqnarray}
    J^\mu = \rho_c v^\mu \,,~~~\partial_\mu J^\mu=0 \,,
    \label{eq:fullu}
\end{eqnarray}
where we define $v^\mu = u^\mu /M$ and again $\rho_c$ has the interpretation of a charge density. 
Now we can consider charges
\begin{eqnarray}
    Q_C = \int d^3\vec x \,C(t, \vec x)\, J^0 \,,
    \label{eq:chargeu}
\end{eqnarray}
for any function $C(t, \vec x)$ that satisfies the equation $u^\mu \partial_\mu C(t, \vec x)=0$.
These are conserved because
\begin{eqnarray}
    \partial_0 Q_C &=&  
    \int d^3\vec x  \,\left(  \partial_0 C(t,\vec x)\, J^0 + C(t,\vec x)\, \partial_0J^0 \right) \nonumber \\
    &=&  \int d^3\vec x  \,\left( u^0 \partial_0 C(t,\vec x)\, \rho_c  - C(t,\vec x)\, \partial_i J^i \right) \nonumber  \\ 
    &=&  \int d^3\vec x  \,\left( -u^i \partial_i C(t,\vec x)\, \rho_c  - C(t,\vec x)\, \partial_i J^i \right)  \nonumber \\
    &=& 
    - \int d^3\vec x  \,\partial_i\left(  C(t,\vec x) J^i \right) =0 \, ,
\end{eqnarray}
assuming sufficient fall-off as usual.

The above generalises to include gravity in a straightforward way. Namely, the theories of interest have a covariantly conserved matter current of the form
\begin{eqnarray}
    J^\mu = \rho_c v^\mu \,,~~~\nabla_\mu J^\mu =  \frac{1}{\sqrt{-g}}\partial_\mu (\sqrt{-g} J^\mu) = 0\,,
    \label{eq:symGR}
\end{eqnarray}
that again implies an infinite number of conserved charges,
\begin{eqnarray}
    Q_C= \int d^3\vec x \sqrt{-g} \,C(t,\vec x)\, J^0 \ ,
    \label{eq:chargeuGR}
\end{eqnarray}
which, for all functions satisfying $u^\mu \partial_\mu C(t,\vec x)=0$, satisfy $\nobreak{\partial_0 Q_C = 0}$.  

With an infinite number of conserved charges, we can expect relativistic fractonic theories to inherit many of the peculiarities of the non-relativistic fractonic theories. In particular, we should expect to see restricted dynamics of the fractonic density $\rho_c$. Let us now turn to a general result of these dynamics.
\newline

\noindent\textit{Fractonic Dust:} The dimensionless vector field $\nobreak{v^\mu=u^\mu/M}$ that appears in the fractonic current $\nobreak{J^\mu=\rho_c v^\mu}$ has the interpretation of a velocity field for the density $\rho_c$. 
This current will contribute to the energy momentum tensor, $T^{\mn}$, a term of the form $A \rho_c  v^\mu v^\nu$, where $A$ is a constant that has dimension of mass. 
When $ u^\mu$ takes its vacuum expectation value (vev) and obeys the geodesic equation $\nobreak{ \bar u^\mu\nabla_\mu \bar u^\nu=0}$, this particular contribution to the energy momentum tensor will be  covariantly conserved on its own: 
\begin{eqnarray}
    \nabla_\mu \left(\rho_c  \bar v^\mu\bar v^\nu\right) = \nabla_\mu \left(\rho_c  \bar v^\mu\right)\bar v^\nu+\rho_c \bar v^\mu\nabla_\mu\bar  v^\nu=0 \,,
\end{eqnarray}
with the first term vanishing by the assumed form of the symmetry current of the theory, and the second term vanishing due to the geodesic equation for $\bar v^\mu=\bar u^\mu/M$. 
 
Thus, the contribution of the form $A \rho_c\bar  v^\mu \bar v^\nu$ to the energy-momentum tensor is that of a pressureless dust. It is separately conserved, with  the above argument applying regardless of whatever interactions the fractons may have. An important application is to be found in the homogeneous and isotropic universe, where  this pressureless dust could play the role of dark matter.

When the field $u^\mu$ deviates from geodesic motion, the contribution to the energy-momentum tensor of the fractonic dust will not conserve on its own, thereby developing pressure. These deviations are described at low energy on general grounds by the analysis of the Goldstone fluctuations of $u^\mu$ around its vev. {These dynamics will in general depend on the details of how Lorentz boosts are broken.} The pressure would naturally avoid problems of caustics associated with an exactly pressureless effective fluid description, see \textit{e.g.}~the discussion in \cite{Arkani-Hamed:2005teg}; these issues could be interesting to study in their own right, given that the microscopic details of fractonic quasiparticles are unusual (see {\it e.g.}~\cite{Seiberg:2020bhn,Seiberg:2020wsg}).
\newline

\noindent\textit{Example---A Fractonic Scalar Field:}
We now turn to a specific example of a theory with a fractonic complex scalar field $\Phi$ and where we choose $u^\mu$ to be a framid. The non-relativistic version of the fractonic scalar  has been studied in \textit{e.g.}~\cite{Pretko:2018jbi,Seiberg:2019vrp}. In that case, the field $\Phi$ transforms not only under the usual global $U(1)$ symmetry $\Phi\to e^{i \alpha} \Phi$ for constant $\alpha$, but under the global symmetries $\Phi  \to e^{i\,C(\vec x)} \Phi $ for arbitrary functions $C(\vec x)$ of spatial coordinates. This symmetry thus precludes the existence of terms in the Lagrangian of the theory that involve spatial derivatives acting on $\Phi$, in distinction to non-fractonic non-relativistic scalar field theories, and leads to the special matter current of the form Eq.~\eqref{eq:fullf}.

Let us now consider the relativistic case. The relativistic version of the above global symmetries are those that act on the complex scalar $\Phi$ as
\begin{equation}
    \Phi  \to e^{i\,C(t,\vec x)} \Phi \,,
\label{eq:symmphi}
\end{equation}
where $C(t,\vec x)$ is any function that satisfies 
\begin{equation}
  u^\mu\partial_\mu C(t,\vec x) = 0 \,.
  \end{equation}

We consider the following effective action involving $\Phi$, $\Phi^*$, $u^\mu$, and their derivatives,
\begin{eqnarray}
   S&=& \int d^4x\sqrt{-g} \bigg( \left(\frac{i}{2}\Phi^* u^\mu \partial_\mu \Phi+h.c.\right)  -   m^2 |\Phi|^2 \nonumber \\ 
    &&~~~~~~~~~~~~~~- \kappa |\Phi|^4  + \lambda(u^\mu u_\mu+ M^2)+ \mathcal{L}_u  \bigg)\,,
\label{eq:action1}
\end{eqnarray}
where $h.c.$ stands for Hermitian conjugate,  $\lambda$ is a Lagrange multiplier to enforce the fixed norm condition, $u^2 = -M^2$, and the term $\mathcal{L}_u$ contains the kinetic terms of the framid $u^\mu$ field 
\begin{eqnarray}
  \mathcal{L}_u &=&- \beta_1 (\nabla_\mu u^\nu)^2  -\beta_2 (\nabla_\mu u^\mu)^2  \nonumber \\ 
  &&- \beta_3  \nabla_\mu u^\nu   \nabla_\nu u^\mu -\frac{\beta_4}{M^2} \dot u^\mu \dot u_\mu
\,,
  \nonumber\\
  &\equiv&   K^{\mu\nu}_{~~~ \sigma\rho } \nabla_\mu u^\sigma \nabla_\nu u^\rho   \ ,
\label{eq:ulag}
\end{eqnarray}
where $\dot u^\mu = u^\nu \nabla_\nu u^\mu$. {$\mathcal {L}_u$} includes all operators that have up to two derivatives acting on $u^\mu$; the $\beta_4$ term contributes at the same order in $1/M$ as the other terms when $u^\mu$ takes its vev. 

 Note the lack of a standard kinetic term $|\partial_\mu \Phi|^2$, which is forbidden by the symmetry Eq.~\eqref{eq:symmphi}. 
Indeed, the appearance of a leading single time derivative term in the action is highly reminiscent of
a non-relativistic theory. In writing down the above effective action we have neglected terms suppressed by the symmetry breaking scale $M$ or by higher powers of derivatives {(such as the operator $|u^\mu\partial_\mu \Phi|^2/M^2$)}, and
we have omitted a term of the form $u^\mu \partial_\mu |\Phi|^2$  that can be eliminated by a redefinition of the phase of $\Phi$ in the analysis that follows.  

It is useful to rewrite the scalar  field in polar variables, $\Phi = e^{i\theta} \phi $, whereupon the fractonic symmetry Eq.~\eqref{eq:symmphi}  acts to shift $\theta \to \theta + C(t,\vec x)$, and to further define the mass dimension two variable $\sigma = \phi^2$. The part of the Lagrangian that involves the scalar field becomes
\begin{eqnarray}
   \mathcal{L}_\Phi &=& \sigma u^\mu\partial_\mu \theta  -  m^2 \sigma - \kappa \sigma^2 \,.
\label{eq:fractonlag}
\end{eqnarray}

In the non-relativistic case, where the immobile fractons act as if they were infinitely heavy particles, it is useful to remove the $m^2$ term from the action by a field redefinition (see {\it e.g.}~\cite{Seiberg:2019vrp}).  However, the relativistic case is different: relativistic fractons are only immobile with respect to the $u^\mu$ field, which can fluctuate around its vev. We will see below how the $m$ parameter enters the energy-momentum tensor and the sound speed of the fluctuations of the fractonic dust.

We now turn to analyse the equations of motion (EoM) arising from this theory.  The conservation of the matter current is expressed in the $\theta$ equation of motion,
\begin{eqnarray}
  \nabla_\mu (\sigma u^\mu) = \nabla_\mu J^\mu = 0 \ .  
    \label{eq:thetaEOM}
\end{eqnarray}

The $\sigma$  equation of motion yields
\begin{eqnarray}
\label{eq,phieom}
 \sigma u^\mu \partial_\mu \theta-m^2\sigma-2\kappa \sigma^2 =0 \ ,
\end{eqnarray}
and the $u^\mu$ equation of motion gives
\begin{eqnarray}
     \sigma  \partial_\mu \theta+2 \lambda u_\mu =  2\nabla_\nu J^{\nu}_{~\mu}  +2 \frac{\beta_4}{M^2} \dot u_\nu \nabla_\mu u^\nu \,,
\label{eq:ueom}
\end{eqnarray} 
where we define $J^{\mu}_{~ \sigma } \equiv K^{\mu\nu}_{~~~ \sigma\rho }  \nabla_\nu u^\rho $. Contracting the framid EoM with $u^\mu$, and using the fixed norm condition that arises  from the EoM of the Lagrange multiplier, we can express the Lagrange multiplier as  
\begin{eqnarray}
  2\lambda &=& \frac{1}{M^2} \left(\sigma  u^\mu \partial_\mu \theta - 2 u^\mu \nabla_\nu J^{\nu}_{~\mu} {-2 \frac{\beta_4}{M^2}\dot u_\nu \dot u^\nu  }  \right)\     \label{eqn:lamgeneral}
 \\
    &=&\frac{1}{M^2} \left(m^2\sigma+2\kappa \sigma^2 - 2 u^\mu \nabla_\nu J^{\nu}_{~\mu}  {-2 \frac{\beta_4}{M^2}\dot u_\nu \dot u^\nu  }  \right)\ . \nonumber
\end{eqnarray}

The energy momentum tensor is defined as
\begin{eqnarray}
    T_\mn \equiv - \frac{2 }{\sqrt{-g}}  \frac{\delta S}{\delta g^\mn} \,,
\label{eq,defEMtensor}
\end{eqnarray}
and, following the convention in \cite{Carroll:2004ai}, treating $g^\mn$ and $u^\mu$ as the independent variables,
it takes the form
\begin{eqnarray}
    T^\mn = 2\lambda u^\mu u^\nu   + g^\mn \left(\mathcal{L}_u+\mathcal{L}_\Phi\right) -\frac{2\delta \mathcal{L}_u}{\delta g_{\mn}} \ .
    \label{eq:emtensor2}
\end{eqnarray} 
We now turn to solving the above system of equations in Minkowski and FRW spacetimes.
\newline

\noindent\textit{Dynamics in Minkowski Spacetime:} To gain intuition into the dynamics of a relativistic fractonic dust, we first consider Minkowski space. We begin by analysing the system when $u^\mu$ takes its vev, $\bar u^\mu=M\delta^\mu_0$, which  has a vanishing current, $\nabla_\nu \bar J^{\nu}_{~\mu} = 0$. 

The conservation of the current in Eq.~\eqref{eq:thetaEOM} gives us, 
\begin{eqnarray}
\partial_0 (  \sigma )=0\implies  \sigma = \bar \sigma (\vec x) \ .
\end{eqnarray}
That is, the fractonic symmetry forces the charge density $\rho_c= \bar\sigma(\vec x) M$ to be a function of only the spatial coordinates $\vec x$.  
We find the Lagrange multiplier from Eq.~\eqref{eqn:lamgeneral},
\begin{eqnarray}
     2\lambda = \frac{1}{M^2}\left(m^2 \bar \sigma(\vec x)  + 2\kappa   \bar\sigma(\vec x)^2\right) \,,
    \label{eq:lambdasolution}
\end{eqnarray}
and from the $u^\mu$ EOM, Eq.~\eqref{eq:ueom}, we have
\begin{eqnarray}
    \bar \sigma(\vec x) \,  \partial_0 \theta -2\lambda M &=& 0  \, ,    \label{eq:bkgsoltheta1} \\
    \bar \sigma(\vec x) \partial_i   \theta &=& 0 \,.
    \label{eq:bkgsoltheta2}
\end{eqnarray}
For non-vanishing $\bar \sigma$, the second of the above equations implies that $\theta = \bar\theta (t)$. This forces a consistency relation that must be satisfied, namely
\begin{eqnarray}
    0 =  \partial_i \partial_0   \bar \theta = \partial_i\left( \frac{2\lambda M}{\bar\sigma(\vec x)}\right) = \frac{1}{M } \partial_i\left( 2\kappa \bar\sigma(\vec x)\right) \,.
\end{eqnarray}
This  requires $\bar\sigma(\vec x ) = \sigma_0$ to be a constant.
That is, there is a constraint that arises from the $u^\mu$ EoM that further forces the fracton charge density to be homogeneous, $\rho_c = M\sigma_0$. 

The energy-momentum tensor, Eq.~\eqref{eq:emtensor2}, takes the form
\begin{eqnarray}
    T^{\mu\nu} 
    &=& \left( m^2 \sigma_0 + 2\kappa \sigma_0^2\right) \bar{v}^\mu  \bar{v}^\nu  + \kappa \sigma_0^2 g^{\mu\nu}  \,, 
\end{eqnarray}
which is of course trivially conserved on this constant solution, but one can anticipate the separation into a pressureless fluid of energy density  $\rho_{dust}= m^2\sigma_0  = \frac{m^2}{M} \rho_c$, and another fluid with equation of state $w=1$. 

Let us pause to reflect upon the role that the constraint played in the above. If $\kappa$ had been tuned to zero, there would in fact be no such constraint, and we would find solutions for arbitrary $\sigma(\vec x)$. When interactions between the fracton particles are turned on, the constraint is active and forces the system into the homogeneous solution  when $u^\mu$ takes its vev.

We can learn more about the low energy dynamics of the fractonic dust by considering small fluctuations around this solution. Working at linear order in perturbations, we write $\nobreak{u^\mu = \bar u^\mu + M  \delta^\mu_{~i} \pi^i}$, $\nobreak{\sigma = \bar\sigma+ \sigma_1}$, and $\nobreak{\theta = \bar\theta +  \theta_1}$. We further decompose the framid  fluctuation $\pi^i$ into a longitudinal and transverse piece, $\nobreak{\pi^i = \pi_L^i + \pi^i_T}$, where
$\pi_L^i = \partial^i \chi$ for a scalar function $\chi$, and  $\nobreak{\partial_i \pi^i_T=0}$. One can check that the first order equations, 
\begin{eqnarray}
   0 &=&   \sigma_0 \partial_i  \pi^i +  \partial_0\sigma_1  \,,\\
    0 &=& M \partial_0  \theta_1 -2\kappa \sigma_1 
    \,,\\
    0&=&  \sigma_0 \partial_i \theta_1 + \frac{1}{M} \left(m^2 \sigma_0 + 2\kappa \sigma_0^2\right) \pi_i - 2\partial_\mu J^{\mu}_{~i} \,,
\end{eqnarray}
lead to wave-like solutions for $\sigma_1$, $\theta_1$, $\pi^i_L$, that satisfy a dispersion relation 
 \begin{eqnarray} 
     -\frac{\kappa \sigma_0^2}{M^2(\beta_1+\beta_2+\beta_3)} \left(k^2 -\frac{\omega^2}{c_\kappa^2} \right)=\omega^2\,\left(k^2 -\frac{ \omega^2 }{c_\beta^2}\right) \,,
 \end{eqnarray}
 where the speeds entering the above are
 \begin{eqnarray}
     c_\kappa^2 = \frac{2\kappa \sigma_0}{m^2+2\kappa \sigma_0} \,,~~~c_\beta^2 = \frac{\beta_1+\beta_2+\beta_3}{\beta_{14}} \,,
 \end{eqnarray}
with {$\beta_{14}=\beta_1-\beta_4$}.
At small {$k\lesssim\sqrt{\sigma_0}m/(M\beta_{123})\equiv k_\star$} the dispersion relation yields a gapped mode with  gap  {$\Delta$, where $\Delta^2=-\kappa\sigma_0^2/(\beta_{14}M^2c_\kappa^2)$}, and a gapless mode with a linear dispersion $\omega^2\simeq c_\kappa^2 k^2$. {Note that a negative gap in general signals an instability on timescales larger than $1/|\Delta|$.} 
As per the discussion of the effective action, we see the parameter $m$ can indeed be interpreted like a mass: the propagation of the lowest energy sound waves in the medium are inversely proportional to it. 
Also as anticipated, the propagation speed of the fluctuations {$\vec \pi_L$} are modified 
as a result of the fracton density being pinned to the framid (for a framid alone these propagate at a speed $c_\beta$, see~\cite{Lim:2004js,Nicolis:2013lma}). One can also check that the dispersion of $\vec \pi_T$ are modified in the fractonic dust: they take a form $\omega^2=\Delta^2+c_T^2\,k^2$  with speed $c_T^2 = \beta_1/\beta_{14}$. 
\newline

\noindent\textit{FRW background:} 
In this section, we discuss the cosmology of the above model on a background FRW metric, $ds^2 = -dt^2 + a^2(t) d\vec x^2 $. We will demonstrate that the pressureless dust component arises as a result of the fractonic symmetry, with a secondary fluid arising from the quartic self-interaction that exhibits kination behavior.

We consider the homogeneous and isotropic solution $ \bar u^\mu = M\delta^{\mu}_0$, which satisfies $\bar u^\mu \nabla_\mu \bar u^\nu=0 $, and again look for solutions for the other fields. The general solution to Eq.~\eqref{eq:thetaEOM} is given by
\begin{eqnarray}
\sigma = \frac{\bar \sigma(\vec x)  }{a(t)^3} \,,
    \label{eq:frwsolutionphi}
\end{eqnarray} 
\textit{i.e.} the only time dependence of $\sigma$ is contained in the factor $1/\sqrt{-g}$. 
One can check that the homogeneous solution $\bar\sigma(\vec x)=\sigma_0$ is in fact enforced by the consistency relation from the $u^\mu$ EoM that we discussed above in flat spacetime.
We  solve Eq.~\eqref{eqn:lamgeneral} to obtain
\begin{eqnarray}
    2\lambda = \frac{m^2 \sigma_0 a^3 + 2 \kappa \sigma_0^2}{M^2 a^6 }    + 6 \beta_2   \dot H- 6 (\beta_1 +\beta_3)  H^2    \ ,
\end{eqnarray}
 where a dot represents a derivative with respect to time, and $H = \dot a/ a$ is the Hubble rate.

The energy momentum tensor defined in Eq.~\eqref{eq:emtensor2} is calculated to be
\begin{eqnarray}
    T^{\mu\nu} &=& \frac{m^2 \sigma_0}{a^3} \bar v^\mu \bar v^\nu + \frac{2\kappa \sigma_0^2}{a^6 } \bar v^\mu \bar v^\nu   +  \frac{\kappa \sigma_0^2}{a^6}g^{\mu\nu} \nonumber\\
    &+& \left(\rho_u+ p_u\right) \bar v^\mu \bar v^\nu + p_u g^\mn \,,
    \label{eq:tmunuFRW}
\end{eqnarray}
where the framid---or Einstein-Aether---fluid has energy density $\rho_u$  
and pressure $p_u$ given by, 
\begin{eqnarray}
&& \rho_u= -3 \gamma M^2 H^2   \ , \nonumber\\
&&p_u\delta^i_{~j} = T_{~~ j}^{i(u)} = M^2 \gamma \delta^i_{~ j} \left(H^2 + \frac{2\ddot a}{a}\right)\ ,
\end{eqnarray}
where $\gamma = \beta_1 +3 \beta_2 + \beta_3  $. 
The 
final two terms in Eq.~\eqref{eq:tmunuFRW} 
act to rescale Newton's constant~\cite{Carroll:2004ai}.  The first term in Eq.~\eqref{eq:tmunuFRW} is conserved by itself, and contributes as a pressureless dust. The two terms proportional to $\kappa$ act like a kination fluid with equation of state $w=1$; see e.g.~\cite{Joyce:1996cp,Ferreira:1997hj} for a discussion of kination cosmology. 

We now consider some basic observational requirements on the parameters of this theory if the fractonic dust were to be a dark matter candidate.
To give the correct fractional energy density of DM today $\Omega_{m,0}\approx 0.26 $, we need $ m^2 \sigma_0  = \rho_{\text{crit}}\ \Omega_{m,0} $, with $\rho_{\text{crit}}$ the critical energy density of the universe, such that
\begin{eqnarray}
  m^2 \sigma_0  \approx 0.5\times  10^{-11}  \text{eV}^4\ . 
\end{eqnarray}

Because the parameter $\sigma_0$ also sets the energy density and pressure of the secondary  fluid, we can  expect constraints arising from the effect a kination fluid has on cosmological evolution, see \textit{e.g.}~\cite{Gouttenoire:2021jhk}.
We can get a basic constraint by first asking at what redshift there is an equal amount of energy in  the dust and the secondary matter fluid. Equating the energy densities, we find this occurs at a redshift $z_i$ where 
\begin{eqnarray}
    \kappa \left(\frac{z_i}{10^{8}}\right)^3 \approx  \left(\frac{m}{\text{keV}}\right)^4 \ .
\end{eqnarray}
Assuming an order one value for $\kappa$, the requirement that the universe be in a radiation dominated phase during Big Bang nucleosynthesis (BBN) would be safely satisfied for $m$ above around a keV.
 The connection of properties of the dark matter to those of the secondary fluid is a novel phenomenological consequence of the fractonic symmetry.

{For the above choice of parameters, requiring the instability timescale to be no shorter than the current Hubble expansion timescale  $|\Delta|\lesssim H_0$ requires $\beta_{14}M^2\sim  M_{pl}^2$.} 
We leave a re-evaluation of how the constraints on framid interactions with gravity (and also with the standard model) may be modified in this model by the coupling of the framid to the fractonic scalar field to future work. Here $M$ represents the cutoff of the framid EFT. While an exploration of UV completions is beyond the scope of the present work, we note UV completions of framids have been studied in the literature, see \textit{e.g.}~\cite{Gripaios:2004ms,Armendariz-Picon:2009kfd}. 

We can check just how pressureless this dust is with example values of the parameter $m\sim\,$keV.  For $\nobreak{k\lesssim k_\star}$,  the sound speed of the low energy fluctuations of the fractonic dust is 
\begin{eqnarray}
    c_\kappa^2=\frac{2\kappa\sigma_0}{m^2+2\kappa \sigma_0}\simeq\frac{2\kappa \rho_{crit}\Omega_{m,0}}{m^4}\simeq \kappa \left(\frac{10^{-3}\,\text{eV}}{m}\right)^4 \,,
\label{eq:speeds}
\end{eqnarray}
which, assuming $\kappa$ of order one, easily avoids large-scale structure constraints on dark matter sound speeds, $c_{dm}^2\lesssim10^{-10}$~\cite{Kunz:2016yqy}. On smaller scales, {$k>k_\star$} the fractonic dust could however exhibit interesting deviations from cold dark matter. 
\newline

\noindent\textit{Discussion:}
We have shown that a pressureless dust in the early universe could be a consequence of a relativistic fractonic symmetry. For the example we gave of a fractonic scalar field, we demonstrated how the pressureless dust arises, and we found that the quartic self-interaction led to a secondary kination fluid. This is not the only interaction that one could consider --- exploring other potentials in this model will lead to different secondary fluids with different phenomenology. These fluids can be seen as a corollary of the existence of a symmetry-enforced dust in the early universe, and could be an interesting signature in a fractonic cosmology. 

An obvious question is, can fractonic dust such as that of the scalar fracton model be a realistic candidate for dark matter, and if so, how to distinguish it? To answer this, one can study the linear perturbations of the fractonic dust on an FRW background, and also consider its non-linear evolution. Our study of the low energy fluctuations about the homogeneous solution in Minkowski space provides some intuition as to how the fractonic dust will behave. One can anticipate that the fractonic dust will develop a sound speed at the order of the perturbations, i.e. have a small pressure, and that it will cluster and behave like dark matter at linear order.  It would be interesting to further study how a fractonic dust collapses and whether it forms halos. Because the dust is tied to a field that breaks Lorentz boost symmetry, there would be no need for halo properties to be boost invariant, which could lead to potential signatures.

Beyond the above considerations, if the fractonic dust is a component of our universe, the question arises about its initial conditions. For example, in order to produce the fractons in the first place one would need to consider breaking some of the fractonic symmetries. The simplest option could be to add explicit breaking terms into the action that break the symmetry considered here down to those of (relativistic versions of) dipole symmetric theories~\cite{Pretko:2018jbi}. One could alternatively consider a possible mechanism in an ultraviolet completion of the $u^\mu$ field,  \textit{e.g.}~along the lines of~\cite{Gripaios:2004ms,Armendariz-Picon:2009kfd} for a framid.
How a fractonic dark matter sector would interact with the visible sector, either through fractonic symmetry conserving operators or through operators that explicitly break it, is another avenue to pursue.

Finally, the fluctuations of a framid field about its equilibrium are unusual as they can violate the null energy condition~\cite{Dubovsky:2005xd,Nicolis:2015sra}. While the equilibrium energy-momentum tensor of the framid remains unmodified after its coupling to the fractonic field, the dynamics of the fluctuations are changed. Irrespective of the dark matter and cosmological application, it will be interesting to study this system further.

\section*{Acknowledgments}
This work is supported by the World Premier International Research Center Initiative (WPI) MEXT, Japan. T.M.\ is supported by JSPS KAKENHI grant JP22K18712.  We thank Elisa Ferreira, Ben Gripaios, Simeon Hellerman, David E.~Kaplan, Justin Khoury, Alex Kusenko,  Mikhail Shaposhnikov, and Tsutomu Yanagida  for lively and helpful discussions.

\bibliography{refs}

\end{document}